\documentclass[11pt]{article}
\usepackage{verbatim,amsmath,eufrak}
\usepackage{hyperref}
\usepackage{cite} % prevents showidx from correct work
\usepackage{makeidx}
\hypersetup{
    bookmarks=true,         % show bookmarks bar?
    unicode=false,          % non-Latin characters in Acrobat's bookmarks
    pdftoolbar=true,        % show Acrobat's toolbar?
    pdfmenubar=true,        % show Acrobat's menu?
    pdffitwindow=false,     % window fit to page when opened
    pdfstartview={FitH},    % fits the width of the page to the window
    pdftitle={My title},    % title
    pdfauthor={Author},     % author
    pdfsubject={Subject},   % subject of the document
    pdfcreator={Creator},   % creator of the document
    pdfproducer={Producer}, % producer of the document
    pdfkeywords={keywords}, % list of keywords
    pdfnewwindow=true,      % links in new window
    colorlinks=false,       % false: boxed links; true: colored links
    linkcolor=red,          % color of internal links
    citecolor=green,        % color of links to bibliography
    filecolor=magenta,      % color of file links
    urlcolor=cyan           % color of external links
}
\renewcommand{\arraystretch}{1.2}
\makeatletter
\newdimen\normalarrayskip              % skip between lines
\newdimen\minarrayskip                 % minimal skip between lines
\normalarrayskip\baselineskip
\minarrayskip\jot
\newif\ifold             \oldtrue            \def\new{\oldfalse}
\def\arraymode{\ifold\relax\else\displaystyle\fi} % mode of array entries
\def\eqnumphantom{\phantom{(\theequation)}}     % right phantom in eqnarray
\def\@arrayskip{\ifold\baselineskip\z@\lineskip\z@
     \else
     \baselineskip\minarrayskip\lineskip2\minarrayskip\fi}
\def\@arrayclassz{\ifcase \@lastchclass \@acolampacol \or
\@ampacol \or \or \or \@addamp \or
   \@acolampacol \or \@firstampfalse \@acol \fi
\edef\@preamble{\@preamble
  \ifcase \@chnum
     \hfil$\relax\arraymode\@sharp$\hfil
     \or $\relax\arraymode\@sharp$\hfil
     \or \hfil$\relax\arraymode\@sharp$\fi}}
\def\@array[#1]#2{\setbox\@arstrutbox=\hbox{\vrule
     height\arraystretch \ht\strutbox
     depth\arraystretch \dp\strutbox
     width\z@}\@mkpream{#2}\edef\@preamble{\halign
\noexpand\@halignto
\bgroup \tabskip\z@ \@arstrut \@preamble \tabskip\z@ \cr}%
\let\@startpbox\@@startpbox \let\@endpbox\@@endpbox
  \if #1t\vtop \else \if#1b\vbox \else \vcenter \fi\fi
  \bgroup \let\par\relax
  \let\@sharp##\let\protect\relax
  \@arrayskip\@preamble}
\def\eqnarray{\stepcounter{equation}%
              \let\@currentlabel=\theequation
              \global\@eqnswtrue
              \global\@eqcnt\z@
              \tabskip\@centering
              \let\\=\@eqncr
 \halign to \displaywidth\bgroup
    \eqnumphantom\@eqnsel\hskip\@centering
    $\displaystyle \tabskip\z@ {##}$%
    \global\@eqcnt\@ne \hskip 2\arraycolsep
         $\displaystyle\arraymode{##}$\hfil
    \global\@eqcnt\tw@ \hskip 2\arraycolsep
         $\displaystyle\tabskip\z@{##}$\hfil
         \tabskip\@centering
    &{##}\tabskip\z@\cr}
\begingroup\ifx\undefined\newsymbol \else\def\input#1 {\endgroup}\fi

\newcounter{app}

\def\app{\setcounter{equation}{0}
\def\theequation{A\Roman{app}.\arabic{equation}}\par
   \addvspace{4ex}
   \@afterindentfalse
  \secdef\@app\@dapp}
\newcommand\@app{\@startsection {app}{1}{0ex}%
                                   {-3.5ex \@plus -1ex \@minus -.2ex}%
                                   {2.3ex \@plus.2ex}%
                                   {\normalfont\Large\bf}}

\def\@dapp#1{%
{\parindent \z@ \raggedright  \bf #1}\par\nobreak}
\def\l@app#1#2{\ifnum \c@tocdepth >\z@
    \addpenalty\@secpenalty
    \addvspace{1.0em \@plus\p@}%
    \setlength\@tempdima{8.5em}%
    \begingroup
      \parindent \z@ \rightskip \@pnumwidth
      \parfillskip -\@pnumwidth
      \leavevmode \bfseries
      \advance\leftskip\@tempdima
      \hskip -\leftskip
      #1\nobreak\hfil \nobreak\hb@xt@\@pnumwidth{\hss #2}\par
    \endgroup\fi}
\newcounter{sapp}[app]

\def\sapp{\def\theequation{A\arabic{app}.\arabic{equation}}\par
   \@afterindentfalse
  \secdef\@sapp\@dsapp}
\newcommand\@sapp{\@startsection{sapp}{2}{\z@}%
                                     {-3.25ex\@plus -1ex \@minus -.2ex}%
                                     {1.5ex \@plus .2ex}%
                                     {\normalfont\large\bfseries}}

\def\@dsapp#1{%
{\parindent \z@ \raggedright  \bf #1}\par\nobreak}
\newcommand{\l@sapp}{\@dottedtocline{2}{1.5em}{3em}}
\def\draft{\oddsidemargin -.5truein
        \def\@oddfoot{\sl preliminary draft \hfil
        \rm\thepage\hfil\sl\today\quad\militarytime}
        \let\@evenfoot\@oddfoot \overfullrule 3pt
        \let\label=\draftlabel
        \let\marginnote=\draftmarginnote
   \def\@eqnnum{(\theequation)\rlap{\kern\marginparsep\tt\@eqnlabel}%
\global\let\@eqnlabel\@vacuum}  }

\def\be{\begin{eqnarray}}
\def\ee{\end{eqnarray}}

\def\p{\partial}

\def\beq{\begin{equation}}
\def\eeq{\end{equation}}
\def\ba{\beq\new\begin{array}{c}}
\def\ea{\end{array}\eeq}
\def\be{\ba}
\def\ee{\ea}

\def\Tr{{\rm Tr}\,}

\def\diag{{\rm diag}\,}

\newfont{\Bbbb}{msbm7 scaled 1\@ptsize00}

\newcommand{\z}{\raise-1pt\hbox{$\mbox{\Bbbb Z}$}}

\newfont{\alef}{msbm10 at 11pt}
\newfont {\goth}{eufm10 at 11pt}
\def\mathbb#1{\hbox{{\alef #1}}}

\let\@@savethanks\thanks
\def\thanks#1{\gdef\thefootnote{\alph{footnote}}\@@savethanks{#1}}

\makeatletter
\g@addto@macro \normalsize {%
 \setlength\abovedisplayskip{14pt plus 3pt minus 3pt}%
 \setlength\belowdisplayskip{14pt plus 3pt minus 3pt}%
  \setlength\abovedisplayshortskip{11pt plus 3pt minus 3pt}%
 \setlength\abovedisplayshortskip{11pt plus 3pt minus 3pt}%
}
\makeatother

\bigskip
\bigskip

\title{
\bigskip
{\bf
Open intersection numbers, matrix models and MKP hierarchy} \vspace{.5cm}}
\author{{\bf A. Alexandrov}\thanks{E-mail:  {\tt alexandrovsash at gmail.com}}
\date{ } \\
{\small {\it Freiburg Institute for Advanced Studies (FRIAS), University of Freiburg, Germany \&}}\\
{\small {\it  Mathematics Institute, University of Freiburg, Germany \&}}\\
{\small {\it ITEP, Moscow, Russia}}\\
}

\begin{document}

\setcounter{footnote}{0}

\setcounter{tocdepth}{3}

\maketitle

\vspace{-8.0cm}

\begin{center}
\hfill ITEP/TH-32/14
\end{center}

\vspace{6.5cm}
%\bigskip
\begin{abstract} 
In this paper we conjecture that the generating function of the intersection numbers on the moduli spaces of Riemann surfaces with boundary, constructed recently by R. Pandharipande, J. Solomon and R. Tessler and extended by A. Buryak, is a tau-function of the KP integrable hierarchy. Moreover, it is given by a simple modification of the Kontsevich matrix integral so that the generating functions of open and closed intersection numbers are described by the MKP integrable hierarchy. Virasoro constraints for the open intersection numbers naturally follow from the matrix integral representation. 

\end{abstract}  

\bigskip

{Keywords: enumerative geometry, matrix models, tau-functions, KP hierarchy, Virasoro constraints}\\

%{\small \bf MSC 2010 Primary: 37K10, 53D45, 81R10, 14N10; Secondary: 81T30.}

\begin{comment}
(	37K10  	Completely integrable systems, integrability tests, bi-Hamiltonian structures, hierarchies (KdV, KP, Toda, etc.)
	14N35  	Gromov-Witten invariants, quantum cohomology, Gopakumar-Vafa invariants, Donaldson-Thomas invariants
   53D45  	Gromov-Witten invariants, quantum cohomology, Frobenius manifolds
		81R10  	Infinite-dimensional groups and algebras motivated by physics, including Virasoro, Kac-Moody, $W$-algebras and other current algebras and their representations 
			81R12  	Relations with integrable systems 
				17B68  	Virasoro and related algebras
				22E70  	Applications of Lie groups to physics; explicit representations
				81T30  	String and superstring theories; other extended objects (e.g., branes))
14N10 Enumerative problems (combinatorial problems)
			\end{comment}

%\bigskip

%\bigskip
\newpage 

%\tableofcontents

\def\thefootnote{\arabic{footnote}}

\section*{Introduction}
\addcontentsline{toc}{section}{Introduction}
\setcounter{equation}{0}

Intersection numbers on the moduli spaces of Riemann surfaces is a challenging and complicated subject of enumerative geometry. While for closed Riemann surfaces an effective description is known for more than twenty years \cite{Konts, Witten}, a similar description of open intersection numbers was not available. Recently, in the paper \cite{PST}  the generating function of open intersection numbers was described by the Virasoro constraints and an infinite hierarchy of PDE's, called there the  ``open KdV hierarchy." This important development makes it possible to apply to the subject the theory of matrix models, a power tool of modern mathematical physics. In this paper we present a simple and natural description of the generating function of the open intersection numbers. 

Namely, let us consider a family of the Kontsevich-Penner models
\be
\tau_N=\det(\Lambda)^N {\mathcal C}^{-1} \int \left[d \Phi\right]\exp\left(-{\Tr\left(\frac{\Phi^3}{3!}-\frac{\Lambda^2 \Phi}{2}+N\log \Phi\right)}\right).
\label{matint}
\ee
From the Kontsevich proof of Witten's conjecture \cite{Konts,Witten} we know that intersection theory on the moduli spaces of closed Riemann surfaces is governed by a representative of this family with $N=0$:
\be
\tau_{KW}=\tau_0,
\ee
which is a tau-function of the integrable KdV hierarchy. The main observation of this work is that $N=1$ case corresponds to the open intersection theory. Namely, the extended generating function (which includes descendants of the boundary points), introduced and studied in \cite{Buryak, Buryak2}, coincides with (\ref{matint}) for $N=1$:
\be
\tau_o=\tau_1.
\ee
Then, from matrix model theory it immediately follows that the extended open generating function is a tau-function of the KP hierarchy. Moreover, the variable $N$ in (\ref{matint}) plays the role of the discrete time. Thus, (\ref{matint}) describes a solution of the modified KP (MKP) hierarchy, and, in addition to the KP (KdV) equations the tau-functions $\tau_o$ and $\tau_{KW}$ satisfy the bilinear identity
\begin{equation}
\oint_{{\infty}} e^{\xi ({\bf t}-{\bf t'},z)}z\,
\tau_o ({\bf t}-[z^{-1}])\tau_{KW}({\bf t'}+[z^{-1}])dz =0.
\end{equation}
We claim that the open KdV hierarchy equations, as well as other PDE's, obtained or conjectured in \cite{PST, Buryak, Buryak2} for the generating function of open intersection numbers follow from the equations of the MKP integrable hierarchy.

Virasoro constraints is a natural property of the matrix integrals. The Virasoro constrains, obtained for the tau-function $\tau_1$, are equivalent to the extended Virasoro constraints, derived in \cite{Buryak2}. An advantage of our version of the Virasoro constraints is that they belong to the symmetry algebra of the integrable hierarchy, thus, they are natural from the point of view of integrability.

The present paper is organized as follows. Section \ref{S1} contains material on the Kontsevich--Witten tau-function. In Section \ref{S2} we establish a relation between the generating function of the open intersection numbers and the matrix model (\ref{matint}) for $N=1$. In Section \ref{S3}
we describe some general properties of the matrix integral (\ref{matint}) and identify the equations of the MKP hierarchy with the equations, obtained in \cite{PST,Buryak,Buryak2}. Section \ref{CONC} is devoted to concluding remarks. For the sake of simplicity in this paper we omit the genus expansion parameter (denoted by $u$ in \cite{PST,Buryak,Buryak2}), since it can be easily restored by rescaling of times.

\section{Kontsevich--Witten tau-function}\label{S1}
The closed intersection theory is governed by the Kontsevich--Witten tau-function. 
Let $\overline{\cal M}_{p;n}$ be the Deligne--Mumford compactification of the moduli space of genus $p$ complex curves $X$ with $n$ marked points $x_1, \dots , x_n$. 
Let us associate with a marked point a line bundle ${\cal L}_i$ whose fiber at a moduli point $(X;x_1,\dots,x_n)$ is the cotangent line to $X$ at $x_i$. 
 Intersection numbers of the first Chern classes of these holomorphic line bundles 
 \be
\int_{\overline{\cal M}_{p;n}}\psi_1^{m_1}\psi_2^{m_2}\dots\psi_n^{m_n}=\langle\sigma_{m_1}\sigma_{m_2}\dots\sigma_{m_n}\rangle
\ee
are rational numbers, not equal to zero only if
\be
\sum^n_{i=1} (m_i-1)=3p-3.
\label{dimcon}
\ee
Their generating function\footnote{We use the variables $t_k$, which are the standard variables of the KP hierarchy. From (\ref{Ksum}) it is clear that they do not coincide with the variables, natural for the intersection theory. The difference between two families of variables is given by the factor $(2m+1)!!$ (see also footnote \ref{foot1}). }
\be\label{Ksum}
F^c\left({\bf t}\right)=\left\langle\exp\left(\sum_{m=0}^\infty (2m+1)!!\, t_{2m+1}\sigma_m\right)\right\rangle
\ee
is given by the Kontsevich--Witten tau-function which is a formal series in odd times $t_{2k+1}$ with rational coefficients:
\be
\tau_{KW}\left({\bf t}\right)=\exp\left(F^c\left({\bf t}\right)\right)\\
=1+\frac{1}{6}\,{t_{{1}}}^{3}+\frac{1}{8}\,t_{{3}}
+{\frac {1}{72}}\,{t_{{1}}}^{6}+{\frac {25}{48}}\,t_{{3}}{t_{{1}}}^{3}+
{\frac {25}{128}}\,{t_{{3}}}^{2}+\frac{5}{8}\,t_{{5}}t_{{1}}
+{\frac {1}{1296}}\,{t_{{1}}}^{9}\\+{\frac {49}{576}}\,{t_{{1}}}^{6}t_{{3
}}+{\frac {1225}{768}}\,{t_{{1}}}^{3}{t_{{3}}}^{2}+{\frac {35}{48}}\,{
t_{{1}}}^{4}t_{{5}}+{\frac {1225}{3072}}\,{t_{{3}}}^{3}+{\frac {245}{
64}}\,t_{{5}}t_{{3}}t_{{1}}+{\frac {35}{16}}\,{t_{{1}}}^{2}t_{{7}}+{
\frac {105}{128}}\,t_{{9}}+\dots
\ee
In the Miwa parametrization it is equal to the asymptotic expansion of the Kontsevich matrix integral \cite{Konts,Witten,Witten2,GN,MM,Unification,Towards,IZK} over the Hermitian matrix $\Phi$:
\be
\tau_{KW}\left(\left[\Lambda\right]\right)=\frac{\displaystyle{\int\left[d \Phi\right]\exp\left(-{\Tr\left(\frac{\Phi^3}{3!}+\frac{\Lambda \Phi^2}{2}\right)}\right)}}{\displaystyle{\int\left[d \Phi\right]\exp\left(-{\Tr\frac{\Lambda \Phi^2}{2}}\right)}}.
\label{matint1}
\ee
This integral depends on the external matrix $\Lambda$, which is assumed to be a positive defined diagonal matrix. 
The times $t_k$ are given by the Miwa transform of this matrix:
\be\label{Miwatimes}
t_{k}=\frac{1}{k}\Tr{\Lambda^{-k}}.
\ee 
All $t_k$ can be considered as independent variables as the size of the matrices tends to infinity and in this limit (\ref{matint1}) gives the Kontsevich--Witten tau-function. After the shift of the integration variable
\be
\Phi=X-\Lambda
\ee
one has
\be
\tau_{KW}\left(\left[\Lambda\right]\right)={\mathcal C}^{-1}\int \left[d X\right]\exp\left(-{\Tr\left(\frac{X^3}{3!}-\frac{\Lambda^2 X}{2}\right)}\right),
\ee
where
\be\label{CC}
{\mathcal C}=e^{\Tr \frac{\Lambda^3}{3}}\int\left[d \Phi\right]\exp\left(-{\Tr\frac{\Lambda \Phi^2}{2}}\right).
\ee

The Harish-Chandra--Itzykson--Zuber formula \cite{HC,IZ} allows us to reduce the r.h.s. of (\ref{matint1}) to the ratio of determinants \cite{Unification,Towards}
\be\label{detfrac}
\tau_{KW}\left(\left[\Lambda\right]\right)=\frac{\det_{i,j=1}^N{\Phi^{KW}_i(\lambda_j)}}{\Delta\left(\lambda\right)}.
\ee
Here $\lambda_j$ are the eigenvalues of the matrix $\Lambda$ and
\be
\Delta(z)=\prod_{i<j}(z_j-z_i)
\ee
is the Vandermonde determinant.
The basis vectors $\Phi^{KW}_i$ are given by the integrals
\be\label{KWbint}
\Phi^{KW}_k(z)=\sqrt{\frac{z}{2\pi}}e^{-\frac{z^3}{3}}\int_{-\infty}^\infty   d y\,y^{k-1} \exp\left(-\frac{y^3}{3!}+\frac{yz^2}{2}\right)\\
=\sqrt{\frac{z}{2\pi}}\int_{-\infty}^\infty   d y\,(y+z)^{k-1} \exp\left(-\frac{y^3}{3!}-\frac{y^2z}{2}\right). 
\ee
The coefficients of the basis vectors can be found explicitly, in particular \cite{IZK,Towards}
\be
\Phi^{KW}_1(z)=\sum_{k=0}^\infty\frac{2^k\,\Gamma\left(3k+\frac{1}{2}\right)}{9^k\,(2k)!\,\Gamma\left(\frac{1}{2}\right)}z^{-3k},\\
\Phi^{KW}_2(z)=-\sum_{k=0}^\infty\frac{6k+1}{6k-1}\frac{2^k\,\Gamma\left(3k+\frac{1}{2}\right)}{9^k\,(2k)!\,\Gamma\left(\frac{1}{2}\right)}z^{1-3k}.
\ee

The first line of (\ref{KWbint}) allows us to find the Kac--Schwarz operators of the KW tau-function \cite{Kac,KS2}. Indeed, we have:
\be\label{recre}
\Phi^{KW}_{k+1}(z)=\sqrt{\frac{z}{2\pi}}e^{-\frac{z^3}{3}} \left(\frac{1}{z}\frac{\p}{\p z}\right)\int_{-\infty}^\infty   d y\,y^{k-1} \exp\left(-\frac{y^3}{3!}+\frac{yz^2}{2}\right)=a_{KW}\,\Phi^{KW}_{k}(z), 
\ee
where
\be\label{CS1}
a_{KW}=\frac{1}{z} \frac{\p}{\p z}+z -\frac{1}{2z^2}.
\ee
Thus, the operator $a_{KW}$ preserves the subspace spanned by the vectors $\Phi^{KW}_i$
\be
a_{KW} \left\{\Phi^{KW}\right\} \subset \left\{\Phi^{KW}\right\}
\ee
and it is the Kac--Schwarz operator (see, e.g., \cite{Enumint} for more details).

To construct another Kac--Schwarz operator we use the identity
\be\label{KWqc}
\left(a_{KW}^2-z^{2}\right)\Phi^{KW}_{1}(z)=0.
\ee
From this identity and the recursion relation (\ref{recre}) it follows that
\be\label{bKW}
z^{2}\Phi_k^{KW}=\Phi_{k+2}^{KW}-2(k-1)\Phi_{k-1}^{KW}.
\ee
Thus, 
\be\label{CS2}
b_{KW}=z^{2}
\ee
is also the Kac--Schwarz operator. The Kac--Schwarz operators (\ref{CS1}) and (\ref{CS2}) satisfy the canonical commutation relation
\be\label{CanKW}
\left[a_{KW},b_{KW}\right]=2
\ee
and generate an algebra of the Kac--Schwarz operators for the KW tau-function.

Given a Kac--Schwarz operator, there is an explicit formula for an operator from the algebra $W_{1+\infty}$ such that the corresponding tau-function is an eigenfunction of this operator \cite{Enumint}.
In particular, the above given Kac--Schwarz operators allow us to construct two infinite series of operators, which annihilate the tau-function. One of them guarantees that the tau-function does not depend on even times 
\be\label{CSsim}
\widehat{J}_{k}^{KW}=\frac{\p}{\p t_{2k}}, \,\,\,\,\,\,\, \quad k\geq 1,
\ee
 so that it is a tau-function of the KdV hierarchy. 
 %From the general properties of the Kac--Schwarz operators it follows that the KW tau-function is an eigenfunction of the operators (\ref{CSsim}). 
 
The second series is given by the Virasoro operators 
\be\label{VirKo}
\widehat{L}_k^{KW}=\frac{1}{2}\widehat{L}_{2k}-\frac{1}{2}\frac{\p}{\p t_{2k+3}}+\frac{1}{16}\delta_{k,0},~~~~~k\geq -1,
\ee
which correspond to the Kac--Schwarz operators
\be\label{VirKS}
l_k^{KW}=-\frac{1}{4}\left((b_{KW})^{k+1}a_{KW}+a_{KW}(b_{KW})^{k+1}\right).
\ee
The operators
\be\label{contin}
\widehat{L}_m=\frac{1}{2} \sum_{a+b=-m}a b t_a t_b+ \sum_{k=1}^\infty k t_k \frac{\p}{\p t_{k+m}}+\frac{1}{2} \sum_{a+b=m} \frac{\p^2}{\p t_a \p t_b}
\ee
in (\ref{VirKo}) generate a family of the Virasoro operators from the $W_{1+\infty}$ algebra of the symmetries of the KP hierarchy.

To find corresponding eigenvalues it is enough to check that the operators (\ref{CSsim}) and (\ref{VirKo}) satisfy the commutation relations:
\be
\left[\widehat{J}_k^{KW},\widehat{J}_m^{KW}\right]=0,~~~~~k,m\geq1\\
\left[\widehat{L}_{k}^{KW},\widehat{J}_m^{KW}\right]=-m\widehat{J}_{k+m}^{KW},~~~~~k\geq-1,~m\geq1,
\\
\left[\widehat{L}_{k}^{KW},\widehat{L}_{m}^{KW}\right]=(k-m)\widehat{L}_{k+m}^{KW},~~~~~~k,m\geq-1.
\ee
Since all generators of the algebra can be obtained as commutators of some other generators, the eigenvalues of all of them are equal to zero:
\be\label{KWfirst}
\widehat{J}_m^{KW} \tau_{KW} =0,~~~m\geq1
\ee
and
\be\label{KWvir}
\widehat{L}_{m}^{KW}\tau_{KW}=0,~~~m\geq-1.
\ee

Then, for any function $Z$ depending only on odd times $t_{2m+1}$, we have 
\be\label{chetnech}
\widehat{\mathcal{L}}_k Z=\left(\widehat{L}_{2k}+\frac{1}{8}\delta_{k,0}\right) Z,~~~~~k\geq-1,
\ee
where the operators
\be
\widehat{\mathcal{L}}_m=\sum_{k=1}^\infty \left(2k+1\right)t_{2k+1}\frac{\p}{\p t_{2k+2m+1}}+\frac{1}{2}\sum_{k=0}^{m-1}\frac{\p^2}{\p t_{2k+1} \p t_{2m-2k-1}}+\frac{t_1^2}{2}\delta_{m,-1}+\frac{1}{8}\delta_{m,0},~~~m\geq -1
\label{VirK}
\ee
constitute the same subalgebra of the Virasoro algebra:
\be
\left[\widehat{\mathcal{L}}_n,\widehat{\mathcal{L}}_m\right]=2(n-m)\widehat{\mathcal{L}}_{n+m},~~~k,m\geq-1.
\ee
Thus, the Virasoro constraints (\ref{KWvir}) are equivalent to the standard Virasoro constraints for the KW tau-function
\be
\widehat{\mathcal{L}}_m \tau_{KW}=\frac{\p}{\p t_{2m+3}}\tau_{KW},~~~m\geq-1,
\label{vir}
\ee
which follow from the invariance of the Kontsevich matrix integral.

\section{Open generating function}\label{S2}

In \cite{PST} the intersection theory on the moduli spaces of Riemann surfaces with boundary was investigated. In particular, open intersection numbers in the genus zero were constructed, and the generalization to all higher genera was conjectured. This conjectural all-genera generating function is uniquely specified by the so called open KdV equations and the Virasoro constraints. In \cite{Buryak} it is proved that the open KdV equations and the corresponding Virasoro constraints give equivalent descriptions of the (conjectural) intersection
theory on the moduli space of Riemann surfaces with boundary. In \cite{Buryak2} the generating function introduced in \cite{PST} was generalized to describe also the descendants on the boundary, and the Virasoro constrains for this conjectural generalized (or extended) generating function were proved. Namely, the generating function of open intersection numbers with descendants\footnote{\label{foot1} The function $F^o$ here is the extended open potential $F^{o, ext}$ of \cite{Buryak2}. Below we continue to use the variables, natural from the point of view of the integrable hierarchies and matrix models. Thus, the variables from \cite{Buryak2} are related to our variables as $t_{k}^B=(2k+1)!!\, t_{2k+1}$, $s_k^B=2^{k+1}(k+1)!\, t_{2k+2}$.} 
\be
\tau_o=\exp(F^{o}+F^c),
\ee
where $F^c=\log(\tau_{KW})$, satisfy the linear equations
\be\label{Virop}
\left(\widehat{\mathcal L}_n+ 2\sum_{k=0}^\infty k\, t_{2k} \frac{\p}{\p t_{2k+2n}} +\frac{3n+3}{2}\frac{\p}{\p t_{2n}}+2 t_2\, \delta_{n,-1}+\frac{3}{2}\delta_{n,0} -\frac{\p}{\p t_{2n+3}}\right) \tau_o =0
\ee
for $n\geq -1$.  These Virasoro operators can not be represented as a linear combination of the operators (\ref{contin}) and the operators $t_k$, $\frac{\p}{\p t_k}$, thus it is obvious that they
do not belong to the $W_{1+\infty}$ symmetry algebra of the KP hierarchy.

According to \cite{Buryak2} the open generating function is related to the KW tau-function by the residue formula
\be\label{resfor}
\tau_o ({\bf t}) =\frac{1}{2\pi i}\oint \frac{d z}{z} D(z)\, \tau_{KW}({\bf t} - \left[z^{-1}\right]) \exp(\xi({\bf t},z)),
\ee
where $\xi({\bf t},z)=\sum_{k=1}^\infty t_k z^{k}$ and we use the standard notation
\be
{\bf t} \pm \left[z^{-1}\right]=\left\{t_1\pm\frac{1}{z},t_2\pm\frac{1}{2z^2},t_3\pm\frac{1}{3z^3},\dots\right\}.
\ee
The series
\be
D(z)=1+ \sum_{k=1} \frac{d_k}{z^{3k}}=1+{\frac {41}{24}}\,{z}^{-3}+{\frac {9241}{1152}}\,{z}^{-6}+{\frac {
5075225}{82944}}\,{z}^{-9}+{\frac {5153008945}{7962624}}\,{z}^{-12}+\dots
\ee
is uniquely defined by the equation
\be
a_{KW} \left(\frac{1}{z}D(z) \right)= \Phi_1^{KW}(z),
\ee
where $a_{KW}$ is the Kac--Schwarz operator for the KW tau-function (\ref{CS1}), thus (\ref{recre}) allows us to identify
\be
\Phi^{KW}_0(z)=\frac{1}{z}D(z).
\ee
 One can easily recover the integral representation for this series. Namely, it is given by the steepest descent expansion of the integral
\be
D(z)=\frac{z^{3/2}}{\sqrt{2\pi}}e^{-\frac{z^3}{3}}\int_{C} d\, y \, y^{-1} \exp\left(-\frac{y^3}{3!}+\frac{y z^2}{2}\right),
\ee
with a properly chosen contour $C$.

Let us consider the Kontsevich matrix integral (\ref{matint1}) with $(M+1)\times(M+1)$ matrix $\Lambda=\diag(y_1,y_2,\dots,y_M,-z)$. Then, in the Miwa variables the relation (\ref{resfor}) yields
\be
\tau_o (\left[Y\right])= \frac{1}{2 \pi i} \oint \frac{d z}{z} D(z) \,\tau_{KW}([\Lambda])\det\left({\frac{Y}{Y-z}}\right), 
\ee
where $Y=\diag(y_1,y_2,\dots,y_M)$. In particular, for $M=1$ we have
\be
\tau_o (\left[y\right])=D(y) \frac{\Phi^{KW}_1(-y)\Phi^{KW}_2(y)-\Phi^{KW}_1(y)\Phi^{KW}_2(-y)}{2y}.
\ee
Since\footnote{This relation is valid for the basis vectors of any KdV tau-function. Indeed, consider a KdV tau-function $\tau({\bf t})$ in the Miwa parametrization with $2\times2$ diagonal matrix $\Lambda=\diag(y,-y)$. In this parametrization all odd times (\ref{Miwatimes}) vanishes 
\be
t_{2k+1}=\frac{1}{2k+1}\left(\frac{1}{y^{2k+1}}-\frac{1}{y^{2k+1}}\right)=0
\ee
and the tau-function is identically equal to $1$. On the other hand, the same tau-function in this parametrization can be represented in terms of the basis vectors as a ratio of determinants of the form (\ref{detfrac})
\be
\tau\left(\left[\Lambda\right]\right)=\frac{\Phi_1(y)\Phi_2(-y)-\Phi_1(-y)\Phi_2(y)}{(-2y)}
\ee
so that for any tau-function independent of even times the basis vectors satisfy
\be
\Phi_1(-y)\Phi_2(y)-\Phi_1(y)\Phi_2(-y)=2y.
\ee
} 
\be
\Phi^{KW}_1(-y)\Phi^{KW}_2(y)-\Phi^{KW}_1(y)\Phi^{KW}_2(-y)=2y,
\ee
we have
\be
\tau_o (\left[y\right])=D(y)=y\, \Phi^{KW}_0(y).
\ee

We conjecture, that the extended generating function of open intersection numbers $\tau_o$ is a KP tau-function, fixed by the set of basis vectors:
\be
\Phi_{j}^{o}(z)=z\Phi_{j-1}^{KW}(z)= \frac{z^{3/2}}{\sqrt{2\pi}}e^{-\frac{z^3}{3}}\int d\, y \, y^{j-2} \exp\left(-\frac{y^3}{3!}+\frac{y z^2}{2}\right) ,\,\,\,\,\,\,j=1,2,3,\dots
\ee
so that it is given by the matrix integral
\be\label{MINT}
\tau_o\left(\left[\Lambda\right]\right)=\mathcal{C}^{-1}\,\det(\Lambda){\int \left[d \Phi\right]\exp\left(-{\Tr\left(\frac{\Phi^3}{3!}-\frac{\Lambda^2 \Phi}{2}+\log \Phi\right)}\right)},
\ee
where ${\mathcal C}$ is given by (\ref{CC}). This matrix integral belongs to the family of the generalized Kontsevich models \cite{Adler,Unification,Towards,GKM,MM,IZK}, and, for $M$ (size of the matrix $\Phi$) large enough, has the following expansion
\be\label{pert}
\tau_o=1+ {\frac {13}{8}}\,t_{{3}}+2\,t_{{1}}t_{{2}}+\frac{1}{6}\,{t_{{1}}}^{
3} \\
+  8\,t_{{6}}+{\frac {1}{72}}\,{t_{{1}}}^{6}+\frac{4}{3}\,{t_{{2}}}^{3}+{\frac {37}{48}}\,t_{{3}}{t_{{1}}}^{3}+\frac{1}{3}\,{t_{{1}}
}^{4}t_{{2}}+2\,{t_{{1}}}^{2}{t_{{2}}}^{2}+{\frac {37}{4}}\,t_{{3}}t_{
{1}}t_{{2}}+4\,t_{{4}}{t_{{1}}}^{2}+8\,t_{{4}}t_{{2}}+{\frac {481}{128
}}\,{t_{{3}}}^{2}+{\frac {65}{8}}\,t_{{5}}t_{{1}} \\
+ {\frac {455}{16}}\,t_{{7}}{t_{{1}}}^{2}+{\frac {61}{576}}\,{t_
{{1}}}^{6}t_{{3}}+{\frac {2257}{768}}\,{t_{{1}}}^{3}{t_{{3}}}^{2}+{
\frac {95}{48}}\,{t_{{1}}}^{4}t_{{5}}+{\frac {7665}{128}}\,t_{{9}}+{
\frac {3965}{64}}\,t_{{5}}t_{{3}}t_{{1}}+{\frac {1}{1296}}\,{t_{{1}}}^
{9}+{\frac {29341}{3072}}\,{t_{{3}}}^{3}\\
+{\frac {14}{9}}\,{t_{{1}}}^{3
}{t_{{2}}}^{3}+\frac{1}{3}\,{t_{{1}}}^{5}{t_{{2}}}^{2}+\frac{1}{36}\,{t_{{1}}}^{7}t_{{
2}}+\frac{8}{3}\,t_{{1}}{t_{{2}}}^{4}+32\,t_{{1}}{t_{{4}}}^{2}+64\,t_{{1}}t_{{
8}}+61\,t_{{6}}t_{{3}}+{\frac {28}{3}}\,t_{{6}}{t_{{1}}}^{3}\\
+60\,t_{{5
}}t_{{4}}+30\,t_{{5}}{t_{{2}}}^{2}+\frac{2}{3}\,t_{{4}}{t_{{1}}}^{5}+{\frac {
61}{6}}\,t_{{3}}{t_{{2}}}^{3}+{\frac {245}{4}}\,t_{{7}}t_{{2}}+64\,t_{
{6}}t_{{1}}t_{{2}}+{\frac {125}{4}}\,t_{{5}}{t_{{1}}}^{2}t_{{2}}+32\,t
_{{4}}t_{{1}}{t_{{2}}}^{2}\\
+{\frac {28}{3}}\,t_{{4}}{t_{{1}}}^{3}t_{{2}
}+61\,t_{{4}}t_{{3}}t_{{2}}+{\frac {61}{2}}\,t_{{4}}t_{{3}}{t_{{1}}}^{
2}+{\frac {2257}{64}}\,{t_{{3}}}^{2}t_{{1}}t_{{2}}+{\frac {61}{24}}\,t
_{{3}}{t_{{1}}}^{4}t_{{2}}+{\frac {61}{4}}\,t_{{3}}{t_{{1}}}^{2}{t_{{2
}}}^{2}+\dots,
\ee
which coincides with the expansion of (\ref{resfor}).

The Kac--Schwarz operator for the tau-function (\ref{MINT}) is 
\be
a_{o}=z\, a_{KW}\, z^{-1}=\frac{1}{z}\frac{\p}{\p z} -\frac{3}{2 z^2}+z, 
\ee
so that
\be\label{reco}
\Phi^o_{k+1}(z)=a_o\Phi^o_{k}(z).
\ee
Let us stress that, contrary to the case of the KW tau-function, this tau-function depends both on odd and even times, since $z^2$ is not a Kac--Schwarz operator anymore:
\be
z^2\Phi_{1}^{o}(z)\notin \left\{\Phi^{o}(z)\right\}.
\ee
Nevertheless, from (\ref{reco}) it immediately follows that the operators
\be
l^o_k=-z^{2k+2}a_o=-z^{2k+2}\left(\frac{1}{z}\frac{\p}{\p z} -\frac{3}{2 z^2}+z\right) 
\ee
for $k\geq-1$ belong to the Kac--Schwarz algebra. The operators $l^o_k$  satisfy the Virasoro commutation relations (with the trivial central charge):
\be
\left[l^o_k,l^o_m\right]=2(k-m)l^0_{k+m}.
\ee
Then, from the general properties of the Kac--Schwarz operators \cite{Enumint} it follows that the tau-function $\tau_o$ is an eigenfunction of the corresponding operators:
\be
\widehat{L}^o_{-1}=\widehat{L}_{-2}-\frac{\p}{\p t_1}+2t_2,\\
\widehat{L}^o_{0}=\widehat{L}_{0}-\frac{\p}{\p t_3}+\frac{1}{8}+\frac{3}{2},\\
\widehat{L}^o_{k}=\widehat{L}_{2k}-\frac{\p}{\p t_{2k+3}}+(k+2)\frac{\p}{\p t_{2k}}, \,\,\,\,k>0,
\ee
where the operators $\widehat{L}_k$ are given by (\ref{contin}).
These operators satisfy the commutation relation of the Virasoro algebra
\be
\left[\widehat{L}^o_k,\widehat{L}^o_m\right]=2(k-m)\widehat{L}^o_{k+m},\,\,\,\,\,\,\,k,m\geq-1.
\ee
From these commutation relations it follows that for $k\geq-1$ the eigenvalues of these operators are equal to zero, thus 
\be\label{Symvir}
\widehat{L}^o_{k}\, \tau_o=0,\,\,\,k\geq-1.
\ee

The Virasoro constraints (\ref{Symvir}) can be reduced to the constraints (\ref{Virop}) with the help of relations
\be\label{simpev}
\frac{\p}{\p t_{2k}}\tau_o=\frac{\p^k}{\p t_2^k}\tau_o
\ee
proved in \cite{Buryak2}. Thus, we see that up to the relations (\ref{simpev}) the tau-function, given by the matrix integral (\ref{MINT}) and an extended generating function of \cite{Buryak2} satisfy the same Virasoro constraints.

The KP hierarchy for the generating function $\tau_o$ which, in particular, immediately follows from the conjectural matrix integral representation (\ref{MINT}), is described by the bilinear identity
\begin{equation}
\oint_{{\infty}} e^{\xi ({\bf t}-{\bf t'},z)}
\,\tau_o ({\bf t}-[z^{-1}])\tau_{o}({\bf t'}+[z^{-1}])dz =0.
\end{equation}
The first non-trivial Hirota equation contained here is
\beq\label{hir3}
(D_{1}^4 +3D_{2}^2 -4D_1 D_3 )\tau_o \cdot \tau_o =0.
\eeq
We use the symbols $D_i$ for the ``Hirota derivatives'' 
defined by
\be\label{MKP1}
P(D) f({\bf t})\cdot g({\bf t}) :=P(\p_X)(f({\bf t}+{\bf X})
g({\bf t}-{\bf X})) |_{{\bf X}=0},
\ee
where $P(D)$ is any polynomial in $D_i$, so that (\ref{hir3}) yields the KP equation in terms of the tau-function
\beq
\tau_o\frac{\p^4\tau_o}{\p t_1 ^4}-4\frac{\p\tau_o}{\p t_1}\frac{ \p^3\tau_o}{\p t_1^3}+3\left(\frac{\p ^2\tau_o}{\p t_1 ^2}\right)^2+
3\tau_o\frac{\p ^2\tau_o}{\p t_2^2}-3\left(\frac{ \p\tau_o}{\p t_2}\right)^2
-4\tau_o\frac{\p^2\tau_o}{\p t_1 \p t_3}+4\frac{\p \tau_o}{\p t_1} \frac{\p \tau_o}{\p t_3}=0.
\eeq
In the next section we will consider a more general integrable structure, equations of which are directly related to the equations derived in \cite{PST, Buryak, Buryak2}.

\section{MKP hierarchy}\label{S3}

Let us consider a family of the Kontsevich-Penner models\cite{Konts,Penner}
\be\label{family}
\tau_N= \det(\Lambda)^N {\mathcal C}^{-1} \displaystyle{\int \left[d \Phi\right]\exp\left(-{\Tr\left(\frac{\Phi^3}{3!}-\frac{\Lambda^2 \Phi}{2}+N\log(\Phi)\right)}\right)},
\ee
which for $N=0$ corresponds to the closed intersections and for $N=1$ according to our conjecture corresponds to the open ones. Here $N$ is the independent parameter, which has nothing to do with the size of the matrices.

Corresponding basis vectors 
\be
\Phi_{j}^{N}(z)={z}^N\Phi_{j-N}^{KW}(z)= \frac{z^{N+1/2}}{\sqrt{2\pi}}e^{-\frac{z^3}{3}}\int d\, y \, y^{j-1-N} \exp\left(-\frac{y^3}{3!}+\frac{y z^2}{2}\right) ,\,\,\,\,\,\,j=1,2,3,\dots
\ee
satisfy the recursive relation
\be
a_N \Phi_j^{N}=\Phi_{j+1}^{N},
\ee
where 
\be
a_{N}=z^N \,a_{KW}\, z^{-N}=\frac{1}{z}\frac{\p}{\p z} -\left(N+\frac{1}{2}\right)\frac{1}{z^2}+z 
\ee
is the Kac--Schwarz operator for $\tau_N$. Thus, from the general relation between the Kac--Schwarz operators and the operators from the algebra $W_{1+\infty}$ it immediately follows that $\tau_N$ satisfies the string equation
\be
\left(\widehat{L}_{-2}-\frac{\p}{\p t_1}+ 2N t_2\right)\tau_N=0.
\ee

Moreover, it is straightforward to check that the operators $z^2a_N$ and $z^4a_N-2(N-1)z^2$ are also the Kac--Schwarz operators so that the tau-function satisfy the equations 
 \be
 \widehat{\mathsf{L}}_k \tau_N =0,\,\,\,\,k=-1,0,1,
 \ee
 where 
 \be
 \widehat{\mathsf{L}}_{-1}=\widehat{L}_{-2}-\frac{\p}{\p t_1}+ 2N t_2,\\
 \widehat{\mathsf{L}}_{0}=\widehat{L}_0-\frac{\p}{\p t_3}+\frac{1}{8}+\frac{3N^2}{2},\\
  \widehat{\mathsf{L}}_{1}=\widehat{L}_{2}-\frac{\p}{\p t_{5}}+3N\frac{\p}{\p t_{2}},
 \ee
 and these three operators satisfy the commutation relations
 \be
 \left[\widehat{\mathsf{L}}_{i},\widehat{\mathsf{L}}_{j}\right]=2(i-j)\widehat{\mathsf{L}}_{i+j}.
 \ee
A complete family of the Virasoro and W-constraints can also be obtained by variation of the matrix integral \cite{GKM,Brezin,MorSur} and is derived in \cite{toap}.
 
The functions of the family (\ref{family}) with different $N$ are related with each other by the differential-difference equations of the KP/Toda type \cite{GKM}. In particular, the tau-functions $\tau_0$ and $\tau_1$ satisfy the MKP integrable hierarchy.\footnote{For more details on MKP hierarchy see, e.g., \cite{JM,AZ} and references therein.} It is given by the bilinear identity 
 \begin{equation}\label{bi1}
\oint_{{\infty}} e^{\xi ({\bf t}-{\bf t'},z)}\, z
\,\tau_o ({\bf t}-[z^{-1}])\tau_{KW}({\bf t'}+[z^{-1}])dz =0
\end{equation}
valid for all ${\bf t}, {\bf t'}$. This bilinear identity is equivalent to an infinite series of PDE's, the simplest of which is
\be\label{MKP11}
\left(D_1^2-D_2\right)\tau_o\cdot\tau_{KW}=0.
\ee
Since $\tau_{KW}=\exp(F^c)$ does not depend on even times, from the definition of the ``Hirota derivatives'' (\ref{MKP1}) it immediately follows that all operators $D_{2k}$ in our case can be substituted by $\frac{\p}{\p t_{2k}}$. Then, equation (\ref{MKP11}) is equivalent to 
\be\label{pde1}
\frac{\p F^o }{\p t_2} =2\frac{\p^2 F^c}{\p t_1^2} +\frac{\p^2 F^o}{\p t_1^2} +\left(\frac{\p F^o}{\p t_1}  \right)^2,
\ee
which was derived in \cite{Buryak}. A combination of this equation and the next equation of the MKP hierarchy
\be
\left(D_1^3-4D_3+3D_1D_2\right)\tau_o\cdot\tau_{KW}=0
\ee
leads to
\be\label{pde2}
\frac{\p F^o}{\p t_3} =\frac{\p F^o}{\p t_1} \frac{\p F^o}{\p t_2} +\frac{\p^2 F^c}{\p t_1^2} \frac{\p F^o}{\p t_1} +\frac{\p^2 F^o}{\p t_1\p t_2} - \frac{1}{2}\frac{\p^3 F^o}{\p t_1^3}.
\ee
This is the first equation of the open KdV hierarchy of \cite{PST}.

On the next level we have two equations
\be
\left(6D_4-8D_1D_3+3D_2^2-D_1^4\right)\tau_o\cdot\tau_{KW}=0,\\
\left(2D_4-D_2^2-D_1^2D_2\right)\tau_o\cdot\tau_{KW}=0,
\ee
from which, in particular, it immediately follows the first equation of (\ref{simpev})
\be
\frac{\p}{\p t_4}\,\tau_o=\frac{\p^2}{\p t_2^2}\,\tau_o.
\ee
We conjecture that other equations of the open KdV hierarchy and other equations of \cite{PST, Buryak,Buryak2} also follow from the bilinear identity (\ref{bi1}) of the MKP hierarchy.

\section{Concluding remarks}\label{CONC}

In this paper we present a description of the open intersection numbers by means of the generalized Kontsevich model. It is more then natural to look for other elements of the modern matrix model theory in this case and to apply these elements to the investigation of the open intersection theory. These elements, in particular, include:
\begin{itemize}

\item Cut-and-join type operator\footnote{ An attempt to describe open intersection numbers in terms of the cut-and-join type operator is made in \cite{Ke}.}

\item Givental decomposition

\item (Quantum) spectral curve 

\item Topological recursion

\end{itemize}
The generating function of open intersection numbers should also describe the model of the open topological string with the simplest possible target-space (a point). It would also be interesting to establish the meaning of the whole family (\ref{family}) from the point of view of enumerative geometry. We also expect that other families of the generalized Kontsevich models should be related to $r$-spin versions of the open intersection numbers. Some of the above mentioned topics are considered in \cite{toap}.

\section*{Acknowledgments}
The author is grateful to A.Morozov for useful discussions and an anonymous referee for many helpful suggestions. This work was supported 
in part by RFBR grant 14-01-00547, NSh-1500.2014.2 and 
by Federal Agency for Science and Innovations of Russian Federation.

\end{document}